\documentclass[a4paper]{revtex4}
\usepackage{graphicx}
\usepackage{fancyhdr}
\usepackage{amsmath}
\newcommand{\lrvec}[1]{\overset{\leftrightarrow}{#1}}
\usepackage{color}

\begin{document}

\title{Tree level unitarity and finiteness of electroweak oblique corrections{\footnote{This report is based on Ref.\cite{Nagai:2014cua}}}}


\author{Ryo Nagai}
\affiliation{Department of Physics, Nagoya University, Nagoya 464-8602, Japan}

\begin{abstract}
We study perturbative unitarity and electroweak oblique corrections in the electroweak symmetry breaking models including an arbitrary number of neutral Higgs bosons. Requiring the perturbative unitarity of the high energy scattering amplitudes of weak gauge bosons and the neutral Higgs bosons at tree level, we obtain a set of conditions among the Higgs coupling strengths (unitarity sum rules).
It is shown that the unitarity sum rules require the tree level $\rho$ parameter to be 1 if there are only neutral Higgs bosons. Moreover, we find that the one-loop finiteness of the electroweak oblique corrections is automatically guaranteed once the unitarity sum rules are imposed among the Higgs coupling strengths. Applying the unitarity sum rules, we obtain severe constraints on the mass of the second lightest neutral Higgs boson and the lightest neutral Higgs (a 125GeV Higgs) coupling strength from the results of the electroweak precision tests as well as the unitarity.
\end{abstract}

\maketitle

\thispagestyle{fancy}


%%%%%%%%%%%%%%%%%%%%%%%%%%%%%%%%%%
\section{Introduction}
In 2012, the Large Hadron Collider (LHC) experiments confirmed the existence of a 125 GeV Higgs boson which interacts with the Standard Model (SM) particles. The current experimental results of the measured couplings of the Higgs boson are consistent with the SM predictions. However, the experimental uncertainties are still large. (For example, the 68\% confidence level (CL) uncertainty of the coupling scale factor for weak gauge bosons is approximately 10\%\cite{Higgscoupling}.) The future LHC experiments with high luminosity and the International Linear Collider experiments are expected to be able to measure the discovered Higgs couplings strengths very precisely. If the discovered Higgs coupling strength with weak gauge bosons turns out to deviate from the SM prediction and there is no new physics beyond the SM, two problems occur at least. The first one is the violation of the perturbative unitarity of the high energy weak gauge bosons' scattering amplitudes. The other one is the appearance of non-renormalizable ultraviolet (UV) divergences in radiative corrections. The deviation of the discovered Higgs coupling strength is therefore evidence for new physics which makes the theory unitary and renormalizable. Can we predict the property of the required new physics if the measured value of the discovered Higgs coupling strength turns out to deviate from the SM prediction? In order to answer this question, we study perturbative unitarity and electroweak oblique corrections in the electroweak symmetry breaking scenarios only with neutral Higgs bosons. We find the precision measurements of the discovered Higgs boson are a great probe for investigating the upper mass bound of the extra neutral Higgs bosons. 
%%%%%%%%%%%%%%%%%%%%%%%%%%%%%%%%%%
\section{Unitarity sum rules and Finiteness of electroweak oblique corrections}
In this section, considering the electroweak symmetry breaking scenarios only with neutral scalar particles (Higgs bosons), we (re)derive two sets of conditions among the Higgs coupling strengths. The first set is known as ``unitarity sum rules'' which are required for the cancelation of the perturbatively unitary violating high energy scattering amplitudes of weak gauge bosons and the neutral Higgs bosons at tree level\cite{Gunion:1990kf}. The other set is needed to guarantee the finiteness of electroweak oblique corrections at one-loop level. We then discuss how the unitarity sum rules are related with the finiteness conditions of the electroweak oblique corrections.

We apply the electroweak chiral Lagrangian technique to describe interactions between the weak gauge bosons and the neutral Higgs bosons in an $SU(2)\times U(1)$ gauge invariant manner. The Lagrangian $\mathcal{L}$ is given by
\begin{align}
\mathcal{L}=\mathcal{L}_{\rm gauge}+\mathcal{L}_\chi+\mathcal{L}_{\rm Higgs},
\end{align}
with $\mathcal{L}_{\rm gauge}$, $\mathcal{L}_{\chi}$, and $\mathcal{L}_{\rm Higgs}$ being the $SU(2)\times U(1)$ gauge Lagrangian, the $SU(2)\times U(1)/U(1)$ non-linear sigma model, and the Higgs Lagrangian, respectively. The $\mathcal{L}_\chi$ is given by
\begin{align}\label{Lchi}
\mathcal{L}_\chi&=\frac{v^2}{4}\mbox{tr}[(D^\mu U)^\dag (D_\mu U)]+\beta\frac{v^2}{4}\mbox{tr}[ U^\dag (D^\mu U)\tau_3]\mbox{tr}[ U^\dag (D_\mu U)\tau_3],
\end{align}
with $U$ being the non-linear sigma model field which includes would-be Nambu-Goldstone bosons. Here $\tau_a~(a=1,2,3)$ are the Pauli matrices. The definition of $U$ and its covariant derivative $D_\mu U$ are given by Ref.\cite{Nagai:2014cua}. The coefficient $\beta$ in Eq.(\ref{Lchi}) is related with the tree level $\rho$ parameter, $\rho_0$, as follows: $\rho_0=(1-2\beta)^{-1}$.

The neutral Higgs bosons ${(\phi^0_n,~n=1,\cdots N_0)}$ are introduced as ``matter'' particles in the chiral Lagrangian. Interactions of these Higgs particles with the weak gauge bosons are described by
\begin{align}
\mathcal{L}^{\rm int}_{\rm Higgs}= &-v\sum_{n=1}^{N_0}\kappa^{\phi^0_n}_{WW}\phi^0_n\mbox{tr}[ U^\dag (D^\mu U)\tau_+]\mbox{tr}[ U^\dag (D_\mu U)\tau_-]\nonumber\\
&-\frac{v}{4}\sum_{n=1}^{N_0}\kappa^{\phi^0_n}_{ZZ}\phi^0_n\mbox{tr}[ U^\dag (D^\mu U)\tau_3]\mbox{tr}[ U^\dag (D_\mu U)\tau_3]\nonumber\\
&-\frac{i}{4}\sum_{n=1}^{N_0}\sum_{m=1}^{N_0}\kappa^{\phi^0_n\phi^0_m}_Z(\phi^0_n\lrvec{\partial}_\mu \phi^{0}_{m} )\mbox{tr}[ U^\dag (D_\mu U)\tau_3]\nonumber\\
&-\frac{1}{2}\sum_{n=1}^{N_0}\sum_{m=1}^{N_0}\kappa^{\phi^0_n\phi^0_m}_{WW}\phi^0_n\phi^0_m\mbox{tr}[ U^\dag (D^\mu U)\tau_+]\mbox{tr}[ U^\dag (D_\mu U)\tau_-]\nonumber\\
&-\frac{1}{8}\sum_{n=1}^{N_0}\sum_{m=1}^{N_0}\kappa^{\phi^0_n\phi^0_m}_{ZZ}\phi^0_n\phi^0_m\mbox{tr}[ U^\dag (D^\mu U)\tau_3]\mbox{tr}[ U^\dag (D_\mu U)\tau_3],
\end{align}
with $\tau^\pm\equiv(\tau^1\pm i\tau^2)/2$ and ${\phi^0_n\lrvec{\partial}_\mu \phi^{0}_{m} \equiv \phi^0_n(\partial_\mu \phi^0_m)-(\partial_\mu \phi^0_n)\phi^0_m}$. Each $\kappa$ denotes the magnitude of the interaction of the neutral Higgs bosons. 

%%%%%%%

We are now ready to investigate perturbative unitarity in the present framework. Evaluating the high energy scattering amplitudes of the longitudinally polarized weak gauge bosons and the neutral Higgs bosons, we can obtain a set of conditions among the Higgs coupling strengths which are needed to cancel the unitarity violating high energy scattering amplitudes. The unitarity sum rules at tree level are given by
\begin{align}
{WW\to WW~\mbox{scattering}}:~~&~~~~~ -4+\frac{3}{\rho_0}+\sum_{n=1}^{N_0}\kappa^{\phi^0_n}_{WW}\kappa^{\phi^0_n}_{WW}=0,\label{sumrule1}\\
{WW\to ZZ~\mbox{scattering}}:~~&~~~~~ \frac{1}{\rho_0}-\rho_0\sum_{n=1}^{N_0}\kappa^{\phi^0_n}_{ZZ}\kappa^{\phi^0_n}_{WW}=0,\label{sumrule2}\\
{WW\to \phi^0_nZ~\mbox{scattering}}:~~&~~~~~ \kappa^{\phi^0_n}_{WW}-\rho_0\kappa^{\phi^0_n}_{ZZ}=0,~~{\mbox{and}}~~\sum_{m=1}^{N_0}\kappa^{\phi^0_n\phi^0_m}_{Z}\kappa^{\phi^0_m}_{WW}=0,\label{sumrule3}\\
{WW\to \phi^0_n\phi^0_m~\mbox{scattering}}:~~&~~~~~ \kappa^{\phi^0_n\phi^0_m}_{WW}-\kappa^{\phi^0_n}_{WW}\kappa^{\phi^0_m}_{WW}=0,~{\mbox{and}}~~\kappa^{\phi^0_n\phi^0_m}_{Z}=0,\label{sumrule4}\\
{ZZ\to \phi^0_n\phi^0_m~\mbox{scattering}}:~~&~~ ~~~\kappa^{\phi^0_n\phi^0_m}_{ZZ}-\rho_0\kappa^{\phi^0_n}_{ZZ}\kappa^{\phi^0_m}_{ZZ}-\sum_{l=1}^{N_0}\kappa^{\phi^0_n\phi^0_l}_Z\kappa^{\phi^0_m\phi^0_l}_Z=0\label{sumrule5}.
\end{align}
Notice that if we combine Eq.(\ref{sumrule1}), Eq.(\ref{sumrule2}), and Eq.(\ref{sumrule3}), we obtain a condition on $\rho_0$,
\begin{align}
\frac{1}{\rho_0}(\rho_0-1)=0.
\end{align}  
Thus the unitarity sum rules require $\rho_0$ to be unity or infinity in any electroweak symmetry breaking model if it only possesses neutral Higgs bosons.

%The unitarity arguments therefore allow us to show, without invoking the custodial symmetry explicitly, the $\rho_0$ should be unity or infinity in any perturbative unitary electroweak symmetry breaking model if it only possesses neutral Higgs bosons.
%%%%%%%

Thanks to the gauge invariance of the non-linear sigma model Lagrangian we use, we can also study radiative corrections in the present framework. Let us focus on electroweak oblique corrections, which can be expressed by $S$, $T$ and $U$ parameters for the model satisfying $\rho_0=1$\cite{Peskin:1990zt}. Since the non-linear sigma model Lagrangian is not renormalizable, non-renormalizable UV divergences appear in the electroweak oblique corrections. We can therefore investigate what conditions are required to cancel these non-renormalizable UV divergences. At one-loop level, the finiteness of the $S$, $T$ and $U$ parameters requires the following conditions among the Higgs coupling strengths:
\begin{align}
&{\mbox{Finiteness condition of the}~S~\mbox{parameter}~}:\nonumber\\
&~~~~~~~~~~~~~~~~~~~~~~~~~~~~~~1-\sum_{n=1}^{N_0}\kappa^{\phi^0_n}_{ZZ}\kappa^{\phi^0_n}_{ZZ}-\frac{1}{2}\sum_{n=1}^{N_0}\sum_{m=1}^{N_0}\kappa^{\phi^0_n\phi^0_m}_Z\kappa^{\phi^0_n\phi^0_m}_Z=0,\label{finiteS}\\
&{\mbox{Finiteness condition of the}~T~\mbox{parameter}~}:\nonumber\\
&~~~~~~~~~~\sum_{n=1}^{N_0}\left[\kappa^{\phi^0_n\phi^0_n}_{WW}-2\kappa^{\phi^0_n}_{WW}\kappa^{\phi^0_n}_{WW}-\kappa^{\phi^0_n\phi^0_n}_{ZZ}+2\kappa^{\phi^0_n}_{ZZ}\kappa^{\phi^0_n}_{ZZ}+\sum_{m=1}^{N_0}\kappa^{\phi^0_n\phi^0_m}_{Z}\kappa^{\phi^0_n\phi^0_m}_{Z}\right]=0,~~\mbox{and}~~\\
\nonumber\\
&~~~~~~~~~~\sum_{n=1}^{N_0}\left(-\kappa^{\phi^0_n\phi^0_n}_{WW}+\kappa^{\phi^0_n}_{WW}\kappa^{\phi^0_n}_{WW}+\kappa^{\phi^0_n\phi^0_n}_{ZZ}-\kappa^{\phi^0_n}_{ZZ}\kappa^{\phi^0_n}_{ZZ}-\sum_{m=1}^{N_0}\kappa^{\phi^0_n\phi^0_m}_{Z}\kappa^{\phi^0_n\phi^0_m}_{Z}\right)\frac{M^2_{\phi^0_n}}{v^2}\nonumber\\
&~~~~~~~~~~~~~~~~~~~~~~~~~~~~~~-\frac{3}{4}g^2\sum_{n=1}^{N_0}(\kappa^{\phi^0_n}_{WW}\kappa^{\phi^0_n}_{WW}-\kappa^{\phi^0_n}_{ZZ}\kappa^{\phi^0_n}_{ZZ})-\frac{3}{4}g^2_Y(1-\sum_{n=1}^{N_0}\kappa^{\phi^0_n}_{ZZ}\kappa^{\phi^0_n}_{ZZ})=0,\\
&{\mbox{Finiteness condition of the}~U~\mbox{parameter}~}:\nonumber\\
&~~~~~~~~~~~~~~~~~~~~\sum_{n=1}^{N_0}(-\kappa^{\phi^0_n}_{WW}\kappa^{\phi^0_n}_{WW}+\kappa^{\phi^0_n}_{ZZ}\kappa^{\phi^0_n}_{ZZ})+\frac{1}{2}\sum_{n=1}^{N_0}\sum_{m=1}^{N_0}\kappa^{\phi^0_n\phi^0_m}_Z\kappa^{\phi^0_n\phi^0_m}_Z=0,\label{finiteU}
\end{align}
with $M_{\phi^0_n}$ being the mass of the neutral Higgs boson $\phi^0_n$. It should be noted that the above conditions Eq.(\ref{finiteS})-Eq.(\ref{finiteU}) are automatically satisfied if the Higgs coupling strengths satisfy the unitarity sum rules Eq.(\ref{sumrule1})-Eq.(\ref{sumrule5}) in the present framework. 

%%%%%%%%%%%%%%%%%%%%%%%%%%%%%%%%%%
\section{Constraints on heavy Higgs bosons}
As discussed in the previous section, the one-loop finiteness of electroweak oblique corrections is ensured by the tree level unitarity sum rules. This allows us to evaluate finite contributions to the tree level scattering amplitudes of longitudinally polarized weak gauge bosons and the one-loop electroweak oblique corrections in any perturbatively unitary model only with neutral Higgs bosons. Both finite contributions can be expressed by the mass of the extra Higgs bosons and the discovered Higgs coupling strengths. Considering constraints on these finite contributions, we obtain the upper mass bound of the extra Higgs bosons as a function of the deviation of the discovered Higgs coupling strength. In this section, we investigate the constraints on the second lightest Higgs boson mass,  $M_H$, by using the perturbatively unitary argument and the result of the electroweak precision tests (EWPTs).

The perturbative unitarity requires the maximum eigenvalue of the $S$-wave transition matrix between $WW$ and $ZZ$ states should be less than one-half\cite{Lee:1977eg}. This condition gives us a constraint on $M_H$ as follows:
\begin{align}\label{unitaritylimit}
M^2_H(1-\kappa^2_V)+M^2_h\kappa^2_V<\frac{16\pi}{5}v^2.
\end{align}
Here $M_h$ denotes the mass of the discovered (the lightest) Higgs boson $h$ and $\kappa_V\equiv\kappa^h_{ZZ}=\kappa^h_{WW}$ denotes the coupling strength of $h$, which satisfies the unitarity sum rules given by Eq.(\ref{sumrule1})-Eq.(\ref{sumrule3}).

On the other hand, using the unitarity sum rules, finite corrections to the $S$ and $T$ parameters for sufficiently the heavy extra Higgs bosons, $M_H\gg v$, can be approximated by
\begin{align}
S&\geq S_H\simeq\frac{1-\kappa^2_V}{12\pi}\left[\ln\frac{M^2_H}{M^2_h}+0.86\right]>0,\\
T&\leq T_H\simeq-\frac{3(1-\kappa^2_V)}{4v^2}\frac{(g^2+g^2_Y)}{g^2g^2_Y}(M^2_Z-M^2_W)\left[\ln\frac{M^2_H}{M^2_h}-1.05\right]<0.
\end{align}
Typical value of $U$ parameter prediction is much smaller than the measured value of the $U$ parameter uncertainty $10^{-2}$, we therefore neglect the $U$ parameter constraint.

\begin{figure}[ht]
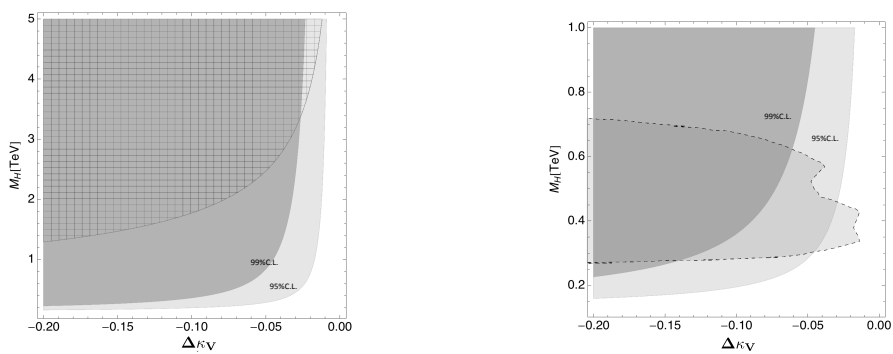

\centering
\includegraphics[width=45mm]{figMHlimit.eps}~~~~~~~~~~~~~~~~~~~~~~~
\includegraphics[width=45mm]{figMHlimit2.eps}
\caption{Constraints on the second lightest Higgs boson mass as function of the deviation of the discovered Higgs coupling.} \label{fig}
\end{figure}

Let us compare the unitary bound and the constraints on the $S$ and $T$ parameters\cite{Baak:2014ora}. These constraints on $M_H$ are summarized in the left-side of Figure \ref{fig}, where $\Delta\kappa_V=\kappa_V-1$ denotes the deviation of the 125GeV Higgs coupling strength. The hatched area is disfavored from the perturbative unitarity. The 95\% and 99\% CL excluded areas from the EWPTs are shown by gray and dark-gray, respectively. We then find, for $\Delta\kappa_V\leq -0.008~(\Delta\kappa_V\leq -0.03)$, the constraints on the $S$ and $T$ parameters at $95\%$ CL ($99\%$ CL) give more stringent bound on $M_H$ than the unitarity bound. Furthermore, we compare these constraints with the results of the LHC direct search for the heavy neutral Higgs boson, shown by the right-side of Figure \ref{fig}. The region surrounded by the dashed contour is excluded by the CMS direct search\cite{kn:CMS-PAS-HIG-13-014}. The results from the LHC direct search gives the strongest bound on $\kappa_V$ for $M_H\simeq 400$GeV, while for the wide range of $M_H$ region the EWPTs have the best sensitivity.

%%%%%%%%%%%%%%%%%%%%%%%%%%%%%%%%%%
\section{Summary}
We consider the electroweak symmetry breaking models including an arbitrary number of neutral Higgs bosons. Using the electroweak chiral Lagrangian technique, interactions between these Higgs bosons and weak gauge bosons can be described in a gauge invariant manner. Thanks to the gauge invariance of our Lagrangian description, we can investigate a relation between the tree level perturbative unitarity of the longitudinally polarized weak gauge bosons' scattering amplitudes and the one-loop finiteness of electroweak oblique corrections. We find the one-loop finiteness of the electroweak oblique corrections is automatically guaranteed by the tree level perturbative unitarity sum rules. Applying the unitarity sum rules, we can test any perturbatively unitary model with neutral Higgs bosons extension by the results of the electroweak precision tests as well as the perturbatively unitary argument. These constraints give us the upper mass bound on the second lightest neutral Higgs boson as a function of the deviation of the discovered Higgs coupling strength. Once the deviation of the discovered Higgs coupling strengths with the weak gauge bosons is experimentally confirmed in future experiments, these results can be used to predict properties of the extra neutral Higgs boson.

\begin{acknowledgments}
Author is grateful to M. Tanabashi and K. Tsumura for fruitful collaborations.
This work is supported by Research Fellowships of the Japan Society for the Promotion of Science (JSPS) for Young Scientists No.26$\cdot$3947.
\end{acknowledgments}
\bigskip % extra skip inserted

\begin{thebibliography}{99} % Use for 10-99 references


%\cite{Nagai:2014cua}
\bibitem{Nagai:2014cua} 
  R.~Nagai, M.~Tanabashi and K.~Tsumura,
  %``Does unitarity imply finiteness of electroweak oblique corrections at one-loop? Constraining extra neutral Higgs bosons,''
  Phys.\ Rev.\ D {\bf 91}, no. 3, 034030 (2015)
 % [arXiv:1409.1709 [hep-ph]].
  %%CITATION = ARXIV:1409.1709;%%



\bibitem{Higgscoupling}
  The ATLAS collaboration,
  %``Measurements of the Higgs boson production and decay rates and coupling strengths using pp collision data at ãs = 7 and 8 TeV in the ATLAS experiment,''
  ATLAS-CONF-2015-007, ATLAS-COM-CONF-2015-011.
  %%CITATION = ATLAS-CONF-2015-007, ATLAS-COM-CONF-2015-011;%%
  %5 citations counted in INSPIRE as of 22 Apr 2015
%\cite{Khachatryan:2014jba}
  V.~Khachatryan {\it et al.}  [CMS Collaboration],
  %``Precise determination of the mass of the Higgs boson and tests of compatibility of its couplings with the standard model predictions using proton collisions at 7 and 8 TeV,''
 % arXiv:1412.8662 [hep-ex].
  %%CITATION = ARXIV:1412.8662;%%
  %59 citations counted in INSPIRE as of 22 Apr 2015




%\cite{Gunion:1990kf}
\bibitem{Gunion:1990kf}
  J.~F.~Gunion, H.~E.~Haber and J.~Wudka,
  %``Sum rules for Higgs bosons,''
  Phys.\ Rev.\ D {\bf 43} (1991) 904.
  %%CITATION = PHRVA,D43,904;%%
  %52 citations counted in INSPIRE as of 07 Aug 2013
%\cite{Grinstein:2007iv}
%\bibitem{Grinstein:2007iv}
  B.~Grinstein and M.~Trott,
  %``A Higgs-Higgs bound state due to new physics at a TeV,''
  Phys.\ Rev.\ D {\bf 76} (2007) 073002
  %[arXiv:0704.1505 [hep-ph]].
  %%CITATION = ARXIV:0704.1505;%%
  %32 citations counted in INSPIRE as of 10 Sep 2014

%\cite{Peskin:1990zt}
\bibitem{Peskin:1990zt}
  M.~E.~Peskin and T.~Takeuchi,
  %``A New constraint on a strongly interacting Higgs sector,''
  Phys.\ Rev.\ Lett.\  {\bf 65} (1990) 964.
  %%CITATION = PRLTA,65,964;%%
  %1380 citations counted in INSPIRE as of 30 Jul 2014


%\cite{Lee:1977eg}
\bibitem{Lee:1977eg}
  B.~W.~Lee, C.~Quigg and H.~B.~Thacker,
  %``Weak Interactions at Very High-Energies: The Role of the Higgs Boson Mass,''
  Phys.\ Rev.\ D {\bf 16} (1977) 1519.
  %%CITATION = PHRVA,D16,1519;%%
  %1623 citations counted in INSPIRE as of 29 Jul 2014

%\cite{Baak:2014ora}
\bibitem{Baak:2014ora}
  M.~Baak {\it et al.}  [Gfitter Group Collaboration],
  %``The global electroweak fit at NNLO and prospects for the LHC and ILC,''
  Eur.\ Phys.\ J.\ C {\bf 74} (2014) 9,  3046
 % [arXiv:1407.3792 [hep-ph]].
  %%CITATION = ARXIV:1407.3792;%%
  %27 citations counted in INSPIRE as of 05 Jan 2015 

\bibitem{kn:CMS-PAS-HIG-13-014}
CMS Collaboration,
%``Search for a heavy Higgs boson in the $H\to ZZ \to 2\ell 2\nu$
%channel in $pp$ collisions at $\sqrt{s}=7$ and 8 TeV'',
Report No. CMS PAS-HIG-13-014.
 
\end{thebibliography}

\end{document}